\documentclass[floatfix,aps,twocolumn,preprintnumbers,amsmath,amssymb,nofootinbib,superscriptaddress,showkeys,showpacs]{revtex4}

\usepackage{graphicx,color}
\usepackage{amsmath,amssymb,bm}
\usepackage{dcolumn}

\def\XXint#1#2#3{{\setbox0=\hbox{$#1{#2#3}{\int}$}
     \vcenter{\hbox{$#2#3$}}\kern-.5\wd0}}

\def\1{\'{\i}}

\begin{document}

\title{Fermi-momentum dependence of relativistic effective mass 
below saturation from 
superscaling of quasielastic electron scattering}

\author{V.L. Martinez-Consentino}\email{victormc@ugr.es} 
\affiliation{Departamento de
  F\'{\i}sica At\'omica, Molecular y Nuclear \\ and Instituto Carlos I
  de F{\'\i}sica Te\'orica y Computacional \\ Universidad de Granada,
  E-18071 Granada, Spain.}

\author{I. Ruiz Simo}\email{ruizsig@ugr.es} \affiliation{Departamento de
  F\'{\i}sica At\'omica, Molecular y Nuclear \\ and Instituto Carlos I
  de F{\'\i}sica Te\'orica y Computacional \\ Universidad de Granada,
  E-18071 Granada, Spain.}

\author{J.E. Amaro}\email{amaro@ugr.es} \affiliation{Departamento de
  F\'{\i}sica At\'omica, Molecular y Nuclear \\ and Instituto Carlos I
  de F{\'\i}sica Te\'orica y Computacional \\ Universidad de Granada,
  E-18071 Granada, Spain.}
  
\author{E. Ruiz
  Arriola}\email{earriola@ugr.es} \affiliation{Departamento de
  F\'{\i}sica At\'omica, Molecular y Nuclear \\ and Instituto Carlos I
  de F{\'\i}sica Te\'orica y Computacional \\ Universidad de Granada,
  E-18071 Granada, Spain.}

\date{\today}

\begin{abstract}
\rule{0ex}{3ex} 

The relativistic effective mass $M^*$, and Fermi momentum, $k_F$, are
important ingredients in the determination of the nuclear equation of
state, but they have rarely been extracted from experimental data
below saturation density where translationally invariant nuclear
matter becomes unstable against clusterization into the existing
atomic nuclei. Using a novel kind of superscaling analysis of the
quasielastic cross section electron scattering data involving a
suitable selection criterion and $^{12}$C as a reference nucleus, the
global scaling properties of the resulting set of data for 21
nuclei ranging from $^2$H to $^{238}$U are then analyzed. We find that
a subset of a third of the about $20000$ data approximately scales to
an universal superscaling function with a more constrained uncertainty
band than just the reference $^{12}$C case and provides $M^*$ as a
function of $k_F$.

\end{abstract}

\pacs{24.10.Jv,25.30.Fj,25.30.Pt,21.30.Fe} 

\keywords{
quasielastic electron scattering, 
relativistic mean field, 
relativistic Fermi gas, effective mass.
}

\maketitle


Since the early days of nuclear physics, the concept of nuclear matter
has provided phenomenological guidance and has stimulated over the
years benchmarking microscopic theoretical solutions of the many-body
nuclear problem under the assumption of translational
invariance. Below the saturation density $n_0 = 0.16$ fm${}^{-3}$,
however, nuclear matter becomes unstable against cluster formation
into the finite atomic nuclei; the equation of state of nuclear matter
has no obvious meaning hindering an extraction of nuclear matter
parameters such as the Fermi momentum or the effective mass from
data. In the present work we undertake such a determination by a
scaling analysis of a large body of inelastic electron-nucleus
scattering data around the quasielastic (QE) peak.

Scaling is a powerful tool for analyzing the response of a complex
system to weakly interacting probes from a variety of complex systems
(see \cite{Sic80,Don99a,Don99b,Ben08} and references therein).  The
probes include electrons, neutrons and neutrinos of different energies
and the targets range from solids and liquids, to atoms, nuclei
and nucleons.  In the case considered here of inclusive electron
scattering from nuclei, scaling has been exploited to analyze a large
amount of QE data for different nuclei in terms of a
phenomenological scaling function extracted from the cross section.
It is a remarkable fact that the reduced cross section of different
nuclei scale to the same function $f(\psi')$ of the scaling variable
$\psi'=\psi'(q,\omega)$, for $\psi'<0$ \cite{Don99a}.  This indicates
an universal behavior of the dynamics inside the system that has been
exploited with success for instance to predict neutrino cross sections
from $(e,e')$ data \cite{Ama04,Meg13}. The study of the longitudinal
and transverse response functions allowed to extract a longitudinal
scaling function $f_L(\psi')$ also for $\psi'>0$ \cite{Mai02}, unlike
the transverse response, which does not scale. To obtain a proper
value of the transverse scaling function $f_T(\psi') \ne f_L(\psi')$
one has to resort to the Relativistic Mean Field (RMF) model, which
reproduces well the $f_L(\psi')$. In the SuSA-v2 model this transverse
scaling function was updated in a fit to $(e,e')$ data, requiring an
additional $q$-dependence, with an explicit scaling
violation \cite{Meg16}, although this is not in full contradiction
with the data.

In recent works \cite{Ama15,Ama17} we have performed a new kind of
scaling analysis of the $^{12}$C QE data \cite{archive2}.
The so-called super scaling analysis with relativistic effective mass
$M^*=m^*_N/m_N$ (SuSAM*) is based on the Fermi gas in RMF as a
starting point, modified by introducing a suitable scaling function
$f^*(\psi^*)$.  The scaling variable $\psi^*$ depends, for each
nucleus, on the Fermi momentum $k_F$ and the relativistic effective
mass $m^*_N$.  This analysis benefits from several advantages related
to the traditional $y$-scaling or the $\psi$-scaling
\cite{Alb88,Don99a,Don99b,Ben08} for the study of the QE
peak. First gauge invariance is preserved by using the effective mass
instead of a separation energy as parameter to describe the center of
the QE peak. Second, it exploits the good properties of RMF
when describing the general features of the QE peak
\cite{Ros80,Ser86,Weh93}, by including the dynamics of the RMF into
the definition of the scaling variable through the effective mass. As
a consequence, the transverse response shows naturally the enhancement
produced by the lower components of the nucleon spinors induced by the
mean field. Third our approach allows to extract an unique scaling
function $f^*(\psi^*)$ for $\psi^*>0$ as well, directly from the cross
section data.

In the present SuSAM* approach we write the
QE cross section in factorized form 
\begin{equation}
\left(\frac{d\sigma}{d\Omega d\omega}\right)_{\rm QE}=
\overline{\left(\frac{d\sigma}{d\Omega}\right)}_{en}
\overline{N_n}(q,\omega)
+
\overline{\left(\frac{d\sigma}{d\Omega}\right)}_{ep}
\overline{N_p}(q,\omega).
\end{equation}
The electron-nucleon cross section averaged over the Fermi sea,
$\overline{\left(\frac{d\sigma}{d\Omega}\right)}_{e n,p}$, for nucleons
with mass $m_N^*$, are obtained from the formalism of
ref. \cite{Ama17}.  These elementary cross sections are multiplied by
the effective number of nucleons that can be excited by the electron
per unit of energy transfer $\omega$, and for given momentum transfer
$q$. For neutrons it is given by
\begin{equation}
\overline{N_n}(q,\omega)= \frac{N\xi_F}{m_N^*\eta_F^3\kappa}f^*(\psi^*)
\end{equation}
and a similar definition for the effective number of protons
$\overline{N_p}$.  Here we use dimensionless variables
$\kappa=q/(2m^*_N)$, $\eta_F=k_F/m^*_N$ and
$\xi_F=(1+\eta_F^2)^{1/2}-1$.  Note that the effective number of
protons (neutrons) is not the true number in the relativistic case,
because its integral over $\omega$ is less than the total number of
protons (neutrons), and a kinematic correction normalization factor
is needed to recover the sum rule.

The above
expressions are exact in the RMF for infinite matter. 
The SuSAM* approach assumes that the scaling function
$f^*(\psi^*)$ is an universal function valid for all nuclei. Here it is
extracted from the
QE $(e,e')$ data of a large amount of nuclei.  It depends on
a single scaling variable defined as the minimum kinetic energy of the
initial nucleon, divided by the Fermi kinetic energy.  All the
kinematics are referred to the RMF, where the nucleon mass is replaced
by the relativistic effective mass $m_N^*$ \cite{Ama17}.

\begin{figure}
\includegraphics[width= 8.5cm]{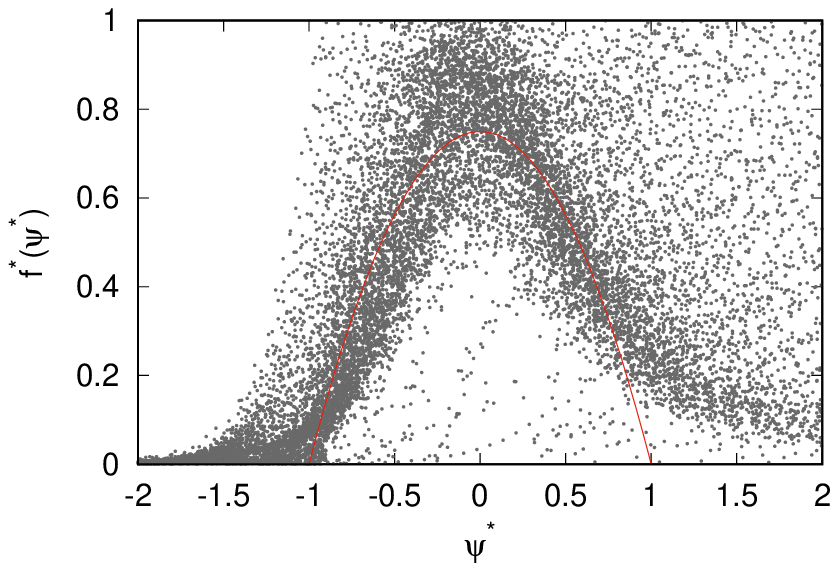}
\includegraphics[width= 8.5cm]{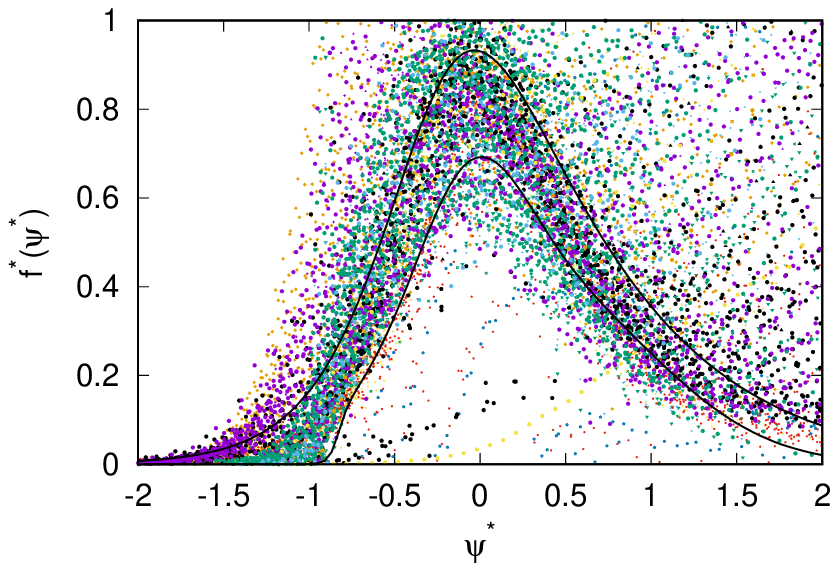}
\caption{Top panel: experimental scaling function $f^*(\psi^*)$
  computed for all the nuclei in the database of ref.
  \cite{Ben08,archive,archive2} compared to the RFG scaling function. 
Bottom panel the
  same data points are compared to the phenomenological band fitted to
  the selected subset of QE data (see text).  }
\end{figure}

In Fig. 1 we show the scaled experimental data used in this work
\cite{archive2,archive}.  
 The
experimental scaling function data in Fig. 1 are computed from the
experimental cross section data by inverting Eq. (1). They have been
plotted against the scaling variable defined as
\begin{equation}
\psi^* = \sqrt{\frac{\epsilon_0-1}{\epsilon_F-1}} {\rm sgn} (\lambda-\tau)
\end{equation}
where
\begin{equation}
\epsilon_0={\rm Max}
\left\{ 
       \kappa\sqrt{1+\frac{1}{\tau}}-\lambda, \epsilon_F-2\lambda
\right\},
\end{equation}
and we use the variables $\lambda = \omega/2m_N^*$, $\tau =
\kappa^2-\lambda^2$, and $\epsilon_F = \sqrt{1+\eta_F^2}$.  Note that
in the traditional $\psi$-scaling variable is a particular case of
these equations for $m_N^*=m_N$.  In both cases the scaling variable
is negative to the left of the QE peak ($\lambda < \tau$)
and positive on the right side.

Note that the data in the top panel of Fig. 1 do not scale in general.
Yet it is remarkable that a large fraction of the data collapse into a
gray band that allows to define a ``quasielastic'' experimental region
with a maximum for $\psi^*=0$. The RFG scaling function is shown as
the parabola in red to compare with.  The data have been scaled with
the values of Fermi momentum and effective mass given in columns 6 and
7 of Table I.  These parameters have been obtained by a $\chi^2$ fit
to the scaling band shown in the bottom panel of Fig. 1.  This band is
parameterized as the sum of two Gaussians modified by a suitable Fermi
function.
\begin{equation}
f^*(\psi^*) = \frac{
 a_3e^{-(\psi^*-a_1)^2/(2a_2^2)}+ b_3e^{-(\psi^*-b_1)^2/(2b_2^2)}}
{1+e^{-\frac{\psi^*-c_1}{c_2}}}
\end{equation}
The parameters of this scaling function are given in Table 2 as well as
the lower, and upper limits of the boundary, defining the uncertainty
band.  The data lying outside the band correspond mainly to
inelastic processes or to low energy excitations of the target
nucleus, both processes breaking naturally scaling not being
quasi-free. The rest of them inside the band correspond approximately
to quasi-free events, the fluctuations generating the band could be
produced by reactions mechanisms beyond the impulse approximation,
including FSI, nuclear correlations, meson-exchange currents,
multi-nucleon knockout, or related processes.

\begin{figure}
\includegraphics[width= 8.5cm]{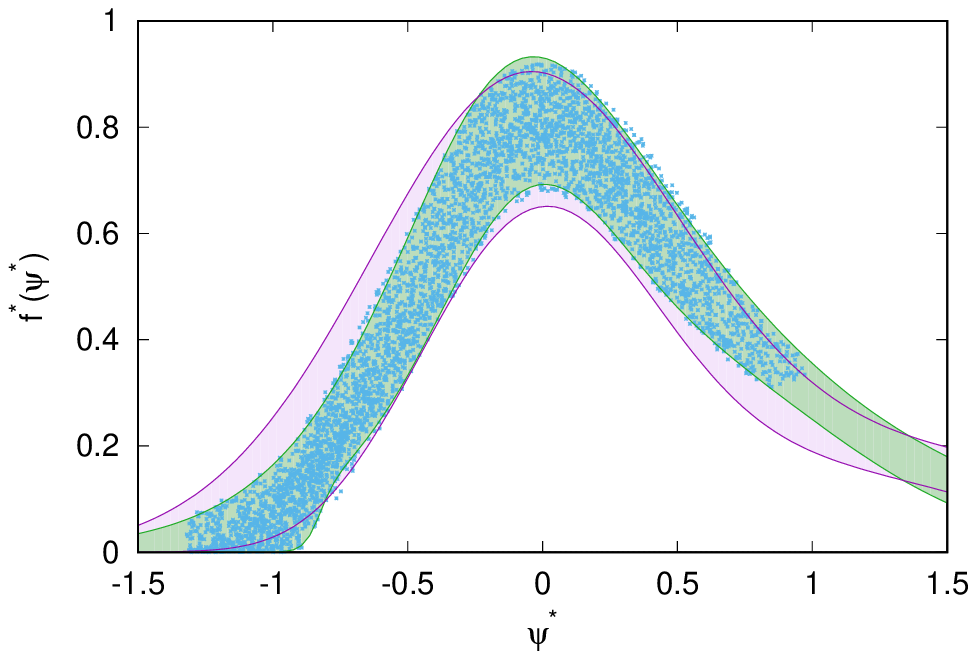}
\includegraphics[width= 8.5cm]{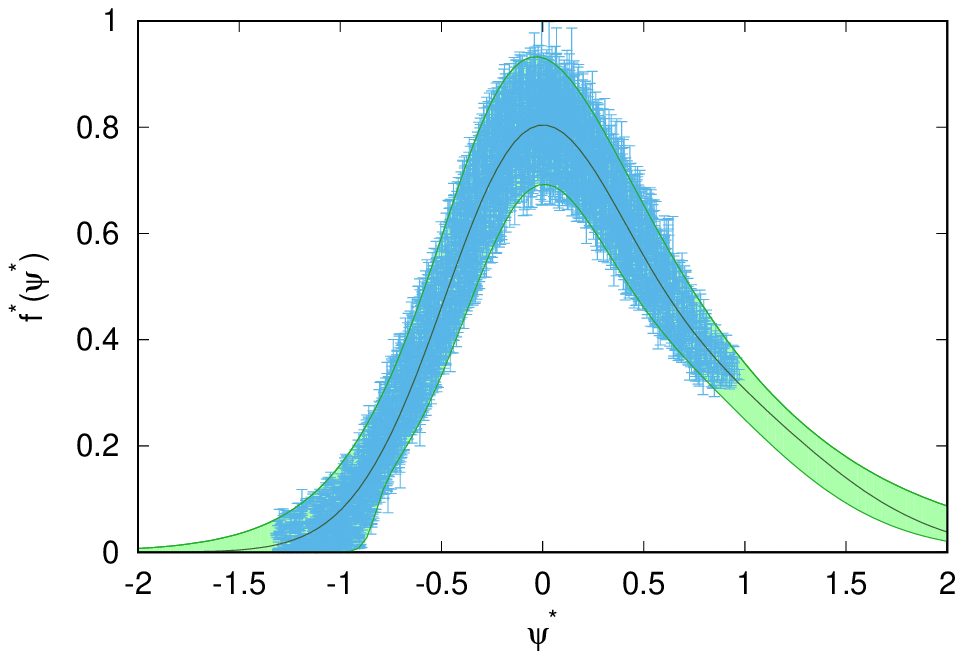}
\caption{Top panel: the starting band in pink of $^{12}$C compared to
  the new fit and to the selected data from all the nuclei. Bottom
  panel: the fitted band compared to the selected data with errors.
   Data are from ref. \cite{Ben08,archive,archive2} }
\end{figure}

To obtain the scaling band is a two-step procedure. 
We have started with the scaling function
already fitted in ref. \cite{Ama17} for $^{12}$C, with parameters
$M^*=0.8$ and $k_F=225$ MeV/c. Our hypothesis is that this starting
band, shown in the top panel of Fig. 2, is approximately valid for every
nucleus. Thus, in this first step we visually tune the
parameters $M^*$ and $k_F$ for each nucleus separately so that the
maximum of the QE peak occurs approximately at $\psi^*=0$
and its width agrees with the $^{12}$C band. The resulting parameters
are shown in the second and third columns of Tab. I.

With these parameters we have scaled all the data selected from those
kinematics where the QE peak is clearly visible. We have
therefore eliminated all the kinematics where only the inelastic or
deep inelastic region are present corresponding to very large momentum
and energy transfer. In the same way we have also discarded the very
low energy kinematics where the peak occurs at $q < 100$ MeV/c and
scaling is violated.  We have only included the nuclei with more than
45 QE data points from Table I.  
With this set of data (not shown) a second
selection has been done by applying the density criterion of
Ref. \cite{Ama15,Ama17}. We have kept only the data 
surrounded by more than 160 points  inside 
a circle of radius $r=0.1$ in the
plot.  

In Fig. 2 we show the resulting set of selected data.  With these data
we have fitted the green band parameterized with Eq. (5), which is compared
to the starting $^{12}$C scaling band of ref. \cite{Ama17}.
The new band turns out to be slightly different and thinner than the starting
 $^{12}$C band.  These differences arise because
the selection procedure has been improved with respect to
ref. \cite{Ama17}. 
 We have verified that the nuclei
with few QE points and not selected are basically inside the fitted
band, hence they do not influence the determination of the band nor
the results on Fig. 2.  In the bottom panel of Fig. 2 we show again
the surviving points with error bars compared to the fitted band.

\begin{table}[t]
\begin{center}
\begin{tabular}{|l|l l|l l|l l| l| l| }
\hline
& \multicolumn{2}{|c|}{Visual fit}
& \multicolumn{2}{|c|}{no. points fit}
& \multicolumn{2}{|c|}{$\chi^2$ fit}
& \multicolumn{2}{|c|}{no. points}\\
\hline 
{Nucleus} & $k_F$ & $M^*$ & $k_F$ & {$M^*$} & $k_F$ & $M^*$ & QE & Total \\
\hline
$^2$H & 80 & 1.00 & 88 & 0.99 & 115 & 0.99 & 426 & 2135\\ \hline
$^3$H & 120 & 0.97 & 142 & 0.99 & 136 & 0.98 & 139 & 540\\ \cline{1-9}
$^3$He & 140 & 0.95 & 147 & 0.96 & 136 & 0.99 & 794 & 2472\\ \cline{1-9}
$^4$He & 160 & 0.90 & 180 & 0.89 & 180 & 0.86 & 803 & 2718\\ \cline{1-9}
$^6$Li & 165 & 0.80 & --- & --- & 175 &  0.77& 23 & 165\\ \cline{1-9}
$^9$Be & 185 & 0.80 & --- & --- & 202 &  0.85 & 16 & 390\\ \cline{1-9}
$^{12}$C & 225 & 0.80 & 226 & 0.82 & 217 & 0.80 & 660 & 2883\\ \cline{1-9}
$^{16}$O & 230 & 0.80 & 259 & 0.84 & 250 & 0.79 & 48 & 126\\ \cline{1-9}
$^{24}$Mg & 235 & 0.75 & --- & --- & 238 & 0.65 & 23 & 34\\ \cline{1-9}
$^{27}$Al & 236 & 0.80 & 258 & 0.78 & 249 & 0.80 & 75 & 628\\ \cline{1-9}
$^{40}$Ca & 240 & 0.73 & 250 & 0.73 & 236 & 0.71 & 616 & 1339\\ \cline{1-9}
$^{48}$Ca & 247 & 0.73 & 242 & 0.75 & 237 & 0.71 & 728 & 1227\\ \cline{1-9}
$^{56}$Fe & 238 & 0.70 & 240 & 0.79 & 241 & 0.70 & 485 & 2429\\ \cline{1-9}
$^{59}$Ni & 235 & 0.67 & --- & --- & 238 & 0.65 & 27 & 37\\ \cline{1-9}
$^{89}$Y & 235 & 0.65 & --- & --- & 224& 0.64 & 27 & 37\\ \cline{1-9}
$^{119}$Sn & 235 & 0.65 & --- & --- & 232 & 0.64 & 24 & 34\\ \cline{1-9}
$^{181}$Ta & 235 & 0.65 & --- & --- & 232 & 0.64 & 24 & 33\\ \cline{1-9}
$^{186}$W & 230 & 0.77 & --- & --- & 226 & 0.76 & 45 & 184\\ \cline{1-9}
$^{197}$Au & 240 & 0.75 & --- & --- & 238 & 0.78 & 30 & 96\\ \cline{1-9}
$^{208}$Pb & 237 & 0.65 & 239 & 0.64 & 233 & 0.56 & 818 & 1714\\ \cline{1-9}
$^{238}$U & 259 & 0.65 & 219 & 0.59 & 219 & 0.51 & 193 & 420\\ 
\hline
\end{tabular}
\end{center}
\caption{Values of the parameters $M^*$ and $k_F$ (in MeV/c) obtained
  from the different fits to the scaling band.}
\label{tabla:final}
\end{table}

\begin{figure}
\includegraphics[width= 8.5cm]{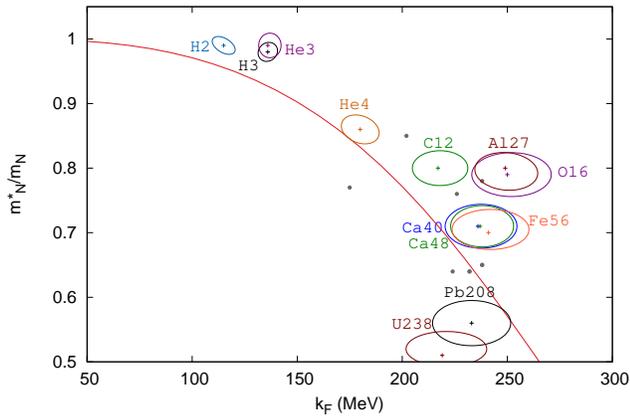}
\caption{
Values of the effective mass represented against the Fermi momentum for all the nuclei considered in the $\chi^2$ fit. The solid line corresponds to the $\sigma-\omega$ model of
Ref. \cite{Ser86}. 
}
\end{figure}

\begin{table*}
\begin{tabular}{ccccccccccc}\hline
   & $a_1$ & $a_2$ & $a_3$ & $b_1$ & $b_2$ & $b_3$ & $c_1$ & $c_2$   \\ \hline
central 
& -0.1335
& 0.4319
& 1.3885
& 0.5741
& 0.6539
& 0.6083
& 0.3405
& 2.2947
\nonumber\\
min
& 0.3075
& -0.6898
& 0.4115
& -0.0647
& -0.3145
& 0.3267
& -0.8362
& 0.0295
\nonumber\\
max
& -7.0719
& -2.4644
& 38.58
& -7.0724
& -2.4595
& 38.58
& -0.2613
& 0.2410
\\ \hline
\end{tabular}
\caption{Parameters of our fit of the phenomenological scaling  
function central value, $f^*(\psi^*)$, and of the lower and upper boundaries (min and max, respectively).} 
\end{table*}

Starting with the parameterized band, Eq. (5), two additional fits of
$M^*$ and $k_F$, have been performed.  First we have maximized the
number of points inside the band. The results are shown in columns 4
and 5 of Tab. 1.  The values are compatible with those estimated
before, specially in the cases where the number of QE data points is
large.  However an unambiguous determination of the effective mass and
Fermi momentum has not been possible by this method 
for some nuclei, due to the small
number of data (their values are not given in the columns 4 and 5 of
Table I). 

To estimate the uncertainty on the parameters, we have
performed a third fitting procedure by minimizing the $\chi^2$
function constructed from the sum of the squared distances to the band
center divided by the band width squared plus the experimental error added in
quadrature. As a result we obtained the scaled data already shown in
Fig. 1. Besides, this fit allows to compute the statistical errors of
the parameters. Only the kinematics where the QE peak is clearly
visible are included in the fit, and only those data where the
momentum transfer is large enough (at least above $q=100$ MeV/c) to
avoid very low energy excitation contributions.

In figure 3 we show the $\chi^2$-fitted values of $M^*$ against the
Fermi momentum, and their $1\sigma$ confidence intervals, represented
by ellipses. With this plot our scaling analysis of the QE peak for
finite nuclei can be related to different realizations (different
$k_F$ values) of nuclear matter below saturation. For comparison we
show the theoretical dependence $M^*(k_F)$ obtained within the RMF in
the $\sigma-\omega$ model of Ref. \cite{Ser86} of nuclear matter. The
points show a similar trend although there is room for
improvements. On the other hand our extracted data provide constraints for
theoretical determinations of the nuclear equation of state.

In the above results the super-scaling hypothesis ---that there is an
universal function $f^*(\psi^*)$ for all nuclei---was verified within
an uncertainty band extracted from the data. However, such an
extraction can be done in multiple ways within the allowed
uncertainty. Thus, in order to test the robustness of our results we
proceed as follows: after the $\chi^2$ fit to the parameterized band,
we generate a second band which is similar to the previous one, and a
new $\chi^2$ fit returns similar values within the present
uncertainties.

To conclude, we have developed a method to obtain the relativistic
effective mass for different values of the Fermi momentum using the
superscaling of the QE data. Our method allows to predict
the $M^*$ values for Fermi momentum below the nuclear matter
saturation point $k_F \sim 270$ MeV/c, corresponding to finite nuclei,
and providing constraints on a empirical determination of the nuclear
equation of state. We have also shown that the $(e,e')$ data allow to
extract a phenomenological scaling function with an uncertainty band
in the QE region.  The scaling function is valid for all
nuclei studied --within their uncertainties-- and points to an
universality in the dynamics of the QE peak. These results
will allow to provide predictions for other nuclei and other reactions
of interest, for example in neutrino scattering from nuclei, of
interest for the neutrino oscillation experiments.

This work is supported by Spanish DGI (grant FIS2014-59386-P) and
Junta de Andalucia (grant FQM225). V.L.M.C. 
acknowledges a contract with
Universidad de Granada funded by Junta de Andalucia and Fondo Social Europeo.
I.R.S. acknowledges support from
the Ministerio de Economia y Competitividad (grant Juan de la
Cierva-Incorporacion).


\end{document}